\documentclass[preprint,showpacs,preprintnumbers,amsmath,amssymb]{revtex4}

\usepackage{graphics}
\newcommand{\intsum}{\hspace{1mm}\int\hspace*{-6mm}\sum}

\begin{document}

\preprint{hep-ph/0405127}
\title{Running Coupling with Minimal Length}

\author{S.~Hossenfelder}
 \affiliation{Department of Physics\\ 
University of Arizona\\
1118 East 4th Street\\ 
Tucson, AZ 85721, USA}
 \email{sabine@physics.arizona.edu}

\date{\today}

\begin{abstract}
In models with large additional dimensions, the GUT scale can be 
lowered to values  accessible by future colliders. Due to modification of the 
loop corrections from particles propagating into the extra dimensions, the 
logarithmic running of the couplings of the Standard Model is turned 
into a power law. These loop-correction are divergent and the standard way
to achieve finiteness is the introduction of a cut-off. The question remains, whether
the results are reliable as they depend on an unphysical parameter.

In this paper, we show that this running of the coupling can be calculated
within a model including the existence of a minimal length scale. The minimal length
acts as a natural regulator and allows us to confirm cut-off computations. 
 \end{abstract}
 

\maketitle

\section{Introduction}
The Standard Model of particle physics yields an extremely precise theory for 
the electroweak and strong interaction. It is renormalizable and physical
observables can be computed, its results  proven by experimental data. 
The Standard Model allowed us to improve our view of nature
in many ways but leaves us with several unsolved problems. 

Among them, the question how to consistently describe quantum effects of gravity is without doubt 
one of the most challenging and exciting problems in physics of this century. When extrapolating 
the strength of the Standard Model interactions by using the renormalization group
equations the three couplings converge. Within the minimal
supersymmetric extension of the Standard Model (MSSM), the couplings meet in one point 
(within the $\alpha_3(M_Z)$ uncertainty) close to $\approx 10^{16}$ GeV \cite{Amaldi:1991cn}.

The study of models with large extra dimensions has recently received a great deal of attention. 
These models,
which are motivated by string theory\cite{dienesundso}, provide us 
with an extension to the Standard Model in which
observables can be computed and predictions for tests beyond the 
Standard Model can be addressed. This in
turn might help us to extract knowledge about the underlying 
theory once we have data to analyze. 
The need to look beyond the Standard Model infected many
experimental groups to search for such SM - violating processes, for a 
summary see e.g. \cite{Azuelos:2002qw}. 

One of the most striking consequences of the large extra dimensions is that unification can occur at
a lowered fundamental scale $M_{\rm f}$, caused by a power law running of the gauge couplings. This modified running of 
the coupling was originally derived by Taylor and Veneziano\cite{Taylor:1988vt} and has been analyzed in
the context of the Standard Model by Dienes, Dudas and Gherghetta\cite{Dienes:1998vh}. The lowered unification scale being one of
the central issues of the models with large extra dimensions, the question of the running coupling
has been addressed in a large number of further works 
\cite{powerlaw,Masip:2000yw,Carone:1999cb,Lorenzana,UVprob,Hebecker,oliver,Quiros}, enlightening the subject 
in many regards.
However, these loop-correction are divergent and the standard way
to achieve finiteness is the introduction of a cut-off $\Lambda$. In this case, the question remains 
whether these results are reliable as they depend on an unphysical parameter.

In this paper we want to demonstrate how the assumption of a minimal length scale $L_{\mathrm f}$ fits in this
scenario naturally. Moreover, the minimal length
removes ambiguities which come along with the cut-off renormalization.

Throughout the whole paper we use the conventions $c= \hbar = 1$, $M_{\rm f} = 1/L_{\rm f}$ and the notation
$\epsilon = L_{\rm f}^2$. Latin indices run over all dimensions.

\section{Large Extra Dimensions}

The recently proposed models of extra dimensions successfully fill 
the gap between theoretical conclusions and experimental possibilities as the extra hidden 
dimensions may have radii large enough to make them
accessible to experiments. Thus, they are an approach towards a phenomenology of grand unified
theories ({\sc GUT}s) at TeV-scale.

There are different ways to build a model of extra dimensional space-time. Here, we want to
mention only the most common ones:
\begin{enumerate}
\item \label{1} The {\sc ADD}-model proposed by Arkani-Hamed, Dimopoulos and Dvali 
\cite{add} adds $d$ extra
spacelike dimensions without curvature, in general each of them compactified to the 
same radius $R$. All Standard Model particles are confined to our brane, while gravitons are 
allowed to propagate freely in the bulk. 
\item \label{2} Within the model of universal extra dimensions ({\sc{UXD}})\cite{dienesundso,uxds,Dienes:1998vh}
all gauge fields (or in some extensions, also fermions) can propagate in the whole 
multi-dimen\-sional spacetime. The extra dimensions are compactified on an orbifold 
to reproduce stand\-ard model gauge degrees of freedom.
\item \label{3} The setting of the model from Randall and Sundrum \cite{rs1,rs2} is a 5-dimensional 
spacetime with
an non-factorizable -- so called warped -- geometry. The solution for the metric is found by analyzing the 
solution of Einsteins field equations with an energy density on our brane, where the 
SM particles live. In the RS 1 model \cite{rs1} the extra dimension is compactified, 
in the RS 2 model \cite{rs2} it is infinite.
\end{enumerate}
It might as well be, that nature chose to realize a mixture of (\ref{1}) and (\ref{2}) or (\ref{2}) and (\ref{3}). 
For a more general review on the subject the reader is referred to \cite{review}.
In the following we will focus on the models (\ref{2}) with $d$ denoting the number of 
this extra dimensions, keeping in mind that there might exist
further dimensions.

In the model of {\sc UXD}s the momentum into the extra dimensions is conserved for gauge boson interactions. \
Therefore, Kaluza-Klein 
excitations can only be produced in pairs; modifications to standard-model processes do not occur
at tree level but arise from loop-contributions. Constraints from electroweak data and collider
experiments thus allow radii to be as large as $1/R \sim$~TeV\cite{moreUXD}.  
Throughout this paper, we fix $1/R=1$~TeV as a representative value.
  
\section{Running Coupling}

In quantum field theory the running of the gauge coupling constants is a consequence of the 
renormalization process, the energy dependence of the coupling constant arising from
loop-contributions to the propagator of the gauge field(s). In a four dimensional spacetime
this contributions are known to be logarithmically divergent $\int {\mathrm d}^4p / p^4 
\sim \int {\mathrm d} p/p \sim \ln p$. In a higher dimensional space-time, divergences get worse. 
As is well known, higher dimensional field theories are non-renormalizable generally.
In this
case one has to introduce a hard cut-off $\Lambda$ in order to render the result finite. The 
existence of extra dimensions then yields a power law explicitly depending on the cut-off parameter
$\Lambda$ which is expected to be in the range of the new fundamental scale. 

There are a vast number 
of publications on this topic \cite{powerlaw}, examining the issue within various classes of unification models and
special regard of one and two step-models \cite{Dienes:1998vh,Carone:1999cb,Lorenzana}. It has been 
investigated\cite{Carone:1999cb} 
how the chosen subset of particles allowed to propagate into the bulk can achieve a more precise
unification point and detailed analysis of two loop
corrections  and threshold effects\cite{Masip:2000yw,oliver} have been given. 

During the last years it has been pointed out, that the
relevant loop corrections suffer from increased UV-sensitivity and that, as a result, no precise statement
can be made about the behavior of the gauge-couplings without first removing the UV-problem (this has e.g.
been mentioned in \cite{UVprob,oliver}). A proposal to this has been made by Hebecker 
and Westphal\cite{Hebecker} by using a
soft breaking of the {\sc GUT}-group symmetry. The fact that the theory is non-renormalizable surely is
due to the fact that is has to be viewed as an effective theory, designed to model a deeper yet not
understood fundamental theory. 

The power law running of the gauge coupling in a higher dimensional spacetime can be explained by
assuming that the $\beta$-function coefficient $b_i$ at an energy $\Lambda$ is proportional to the number of active 
flavors, meaning in this context the number of KK-modes with excitation energies below $\Lambda$. In this
case on finds  
\begin{eqnarray} \label{coeff}
b_i \sim \Omega_d  \left( \Lambda R \right)^d \quad,
\end{eqnarray}  
with $\Omega_d$ being the Volume of the $d$-dimensional sphere
\begin{eqnarray}
\Omega_d = \frac{\pi^{(d/2)}}{\Gamma(1+d/2)} \quad.
\end{eqnarray}  

This dependence on the energy scale is also justified by hard cut-off computations. Introducing an
infrared cut-off $\mu_0$ as well as an ultraviolet cutoff $\Lambda$, the behavior of the
one-loop corrections can be estimated as
\begin{eqnarray} \label{uvir}
\int_{\mu_0}^{\Lambda} \frac{{\mathrm d}^{d+4} p}{p^4} \sim  \Lambda^d - \mu_0^d \quad.
\end{eqnarray}  
Performing this
calculations, one is faced with the problem that the result depends explicitly on the cut-off $\Lambda$. 
This forces one to interpret the cut-off as the renormalization scale $\mu$, giving rise to one-loop-corrected
values of the gauge coupling $\alpha_i(\Lambda)$ as functions of the value of this cut-off parameter.  In
many cases in quantum field theories this cut-off dependence is identical to the scale dependence which can
be computed using reliable renormalization schemes that do not depend on the regulator, e.g. dimensional
regularization.\footnote{It should be mentioned, that it is nonetheless possible to perform a dimensional regularization in
the sense, that it is possible to capture the infinities in a $\Gamma$-function since is has no poles when $d$ 
is not an integer. However, to use this renormalization scheme one needs to introduce a mass-scale to assure
the gauge couplings have the right power. For $d>0$ the result depends explicitly on this mass-scale which might
or might not agree with $M_f$ and thus does not solve the problem.}

In particular, there remain several ambiguities using the cut-off formalism. The first problem at hand is whether
the cut-off $\Lambda$ agrees with the regularization scale $\mu$. Further, the use of a cut-off on the KK-tower immediately raises the question 
for the threshold of the modes and how they are correctly added to the tower. Especially regarding the first mode,
when using the above 
arguments, below the energy $1/R$ there are no
excitations of KK-modes at all. The value $1/R$ thus acts essentially as an infrared cut-off. The higher dimensional theory is 
matched to the four-dimensional logarithmic running at this infrared cut-off. It is unclear within this procedure in which way the crossing of 
the thresholds is performed best and whether the matching point to the theory on the brane is chosen correctly. Since the 
value of the matching point is the onset of the power law-running, its value is crucial for the value of the unification scale.

Further, besides all educated arguments, the constant for the coefficient in (\ref{coeff}) 
finally has to be fixed by hand. This modifies the
slope of the running once the threshold is crossed. All of these problems do not
alter the main point {that} the coupling constants get power law corrections and that they unify at a lowered
scale. But they are unsatisfactory from a theoretical point of view and do not allow us to make 
predictions.
 
As the minimal length we introduce modifies the measure of the momentum space in the ultraviolet region,
the troublesome loop contributions get finite. The minimal length acts as a natural regulator, but in contrast to
computations using cut-off regularization techniques, we expect the result to depend on the new 
parameter as it is an order parameter for physics beyond the Standard Model.

\section{Minimal Length}

\subsection{General Motivation}

Even if a full description of quantum gravity is not yet available, there
are some general features that seem to go hand in hand with all promising candidates
for such a theory. One of them is the need for a higher dimensional spacetime;
another is the existence of a minimal length scale. As the success of string theory arises
from the fact that interactions are spread out on the world-sheet and do no longer take
place at one singular point, the finite extension of the string has to become important at
small distances or high energies, respectively. Now that we are discussing the possibility of
a lowered fundamental scale, we want to examine the modifications arising from this, as they
might get observable soon. If we do so, we should clearly take into account the minimal length
effects.

In perturbative string theory \cite{Gross:1987ar,Amati:1988tn},
the feature of a fundamental minimal length scale arises from the fact that strings can not probe 
distances smaller than the string scale. If the energy of a string reaches this scale $M_s=\sqrt{\alpha'}$, 
excitations of the 
string can occur and increase its extension \cite{Witten:fz}. In particular,
an examination if the spacetime picture of high-energy string scattering shows, that 
the extension of the string grows proportional to its energy\cite{Gross:1987ar} in every order
of perturbation theory. 
Due to this, uncertainty in position measurement can never become arbitrarily small.
For a review, see \cite{Garay:1994en,Kempf:1998gk}.

In this paper we will implement both of these phenomenologically motivated issues of string theory into 
quantum field theory: the extra dimensions and the minimal length. We do not aim to derive them from a 
fully consistent theory of first principles. Instead, we will 
analyze the consequences for the running coupling and ask what conclusions might be drawn 
 for the underlying theory. 

\subsection{Minimal Length in Quantum Mechanics}

Naturally, the minimum length uncertainty is related to a modification 
of the standard commutation relations between position and 
momentum \cite{Kempf:1994su,Kempf:1996nk}. With the Planck scale as high as $10^{16}$~TeV, applications of this
are of high interest mainly for quantum fluctuations in the early universe and for 
inflation processes and have been examined closely\cite{tanh,gup}. 

There are several approaches how to deal 
with the generalization of the relation between momentum and wave vector, 
see e.g.\cite{miniapp}. 
To incorporate the notion of a minimal length into ordinary quantum field theory we will apply a
simple model which has been worked out in detail in\cite{Hossenfelder:2003jz}. 

We assume, no matter how 
much we increase the momentum $p$ of a particle, we can never
decrease its wavelength below some minimal length $L_{\mathrm f}$ or, equivalently, we can never increase
its wave-vector $k$ above $M_{\mathrm f}$. Thus, the relation between the 
momentum $p$ and the wave vector $k$ is no longer linear $p=k$ but a 
function\footnote{Note, that this is similar to introducing an energy 
dependence of Planck's constant $\hbar$.} $k=k(p)$.
 This function $k(p)$ has to fulfill the following properties:

\begin{enumerate}
\item For energies much smaller than the new scale we reproduce the linear relation: 
for $p \ll M_{\mathrm f}$ we have $p \approx k$ \label{limitsmallp}
\item It is an odd function (because of parity) and $k$ is collinear to $p$ (see also Fig. \ref{fig5}).
\item The function asymptotically approaches the upper bound $M_{\mathrm f}$. \label{upperbound}
\end{enumerate} 

We will assume, that $L_{\rm f} \ll R$, so
that the spacing of the Kaluza-Klein excitations compared to energy scales $M_{\rm f}$ 
becomes almost continuous and we can use the integral form.

Lorentz-covariance is not added to the above list, as the proposed model can not provide conservation
of this symmetry. This is easy to see if we imagine an observer who is boosted relative to the minimal length.
He then would observe a contracted minimal length which would be even smaller than the minimal length. To
resolve this problem it might be inevitable to modify the Lorentz-transformation. Several attempts to construct
such transformations have been made\cite{Amelino-Camelia:2002wr} but no clear answers have been given yet. Therefore we will 
assume
$p$ is a Lorentz vector, aim to express all quantities in terms of $p$ and otherwise have to cope with a lack
of Lorentz-covariance in $k$-space. One might think of constructing a covariant relation, but since the only 
covariant
quantity available is $p^2$ and thus a constant\footnote{At least on-shell.} which is fixed by 
(\ref{limitsmallp}) we had no
upper bound (\ref{upperbound}).

A relation fulfilling the above properties might be put in the form
\begin{eqnarray}
k_\mu = \hat{e}_\mu \xi (p_{\rm e}) \quad, \label{defxi}
\end{eqnarray}
where the index 'e' denotes the euclidean norm and $\hat{e}_\mu$ is the unit vector in $\mu$-direction.  
We will specify the exact form later on (see end of
this section).

The quantization of these relations is straightforward and follows the usual procedure. 
The commutators between the corresponding operators $\hat{k}$ and $\hat{x}$ 
remain in the standard form. 
Using the well known commutation relations 
\begin{eqnarray} \label{CommXK}
[\hat x_i,\hat k_j]={\mathrm i } \delta_{ij}\quad
\end{eqnarray}
and inserting the functional relation between the
wave vector and the momentum then yields the modified commutator for the momentum 
\begin{eqnarray} \label{CommXP}
[\,\hat{x}_i,\hat{p}_j]&=& + {\rm i} \frac{\partial p_i}{\partial k_j} \quad.
\end{eqnarray} 
This results in the generalized uncertainty relation
\begin{eqnarray} \label{gu}
\Delta p_i \Delta x_j \geq \frac{1}{2}  \Bigg| \left\langle \frac{\partial p_i}{\partial k_j} 
\right\rangle \Bigg| \quad, 
\end{eqnarray}
which reflects the fact that by construction it is not possible to resolve space-time distances
arbitrarily well. Since $p(k)$ gets asymptotically constant its derivative $\partial p/ \partial k$
drops to zero and the uncertainty in (\ref{gu}) increases for high energies. 
The behavior of our particles thus agrees with those of the strings found by Gross as mentioned above.

The form of the new operator $\hat{p}_i$ is most easily analyzed when we expand the inverted 
relation $p(k)$ in a power-series with coefficients $a_n$. E.g. in the one dimensional case 
suppose we have the series
\begin{eqnarray} 
p_x = k_x + \sum_{{n \geq 1}}  a_n k_x^{2n+1}   \quad. 
\end{eqnarray}
It can then be seen that in position representation the momentum operator takes the form
\begin{eqnarray} 
\hat{p}_x= - {\rm i} \partial_x + \sum_{{n \geq 1}} a_n (- {\rm i})^{2n+1} \partial^{2n+1}_x \quad.
\end{eqnarray}

Since $k=k(p)$ we have for the eigenvectors $\hat{p}(\hat{k})\vert k \rangle= p(k)\vert k \rangle$ and 
so $\vert k \rangle \propto \vert p(k) \rangle$. We could now add that both sets 
of eigenvectors have to be a complete orthonormal system and 
therefore $\langle k' \vert k \rangle = \delta(k-k')$, 
$\langle p' \vert p \rangle = \delta(p-p')$. 
This seems to be a reasonable choice at 
first sight, since  $\vert k \rangle$ is known from the low energy regime. 
Unfortunately, now the normalization of the states is different 
because $k$ is restricted to the Brillouin zone
$-1/L_{\mathrm f}$ to $1/L_{\mathrm f}$. 

To avoid the need to recalculate normalization factors, we  
choose the $\vert p(k) \rangle$ to be identical to 
the $\vert k \rangle$. Following the proposal of \cite{Kempf:1994su} this yields then
a modification of the measure in momentum space.

To make this point more clearly, especially in the presence of compactified extra dimensions,
let ${x}$ be the uncompactified coordinates on our brane and ${y}$ the coordinates in the
direction of the compactified extra dimensions. Since each of the latter is compactified on the same radius $R$, we
have for the $d$-dimensional volume ${\mathrm{Vol}}_d(y)$ of the extra dimensions 
\begin{eqnarray} 
{\mathrm{Vol}}_d(y) = (2 \pi R)^d \quad.
\end{eqnarray}
In addition to this, the volume of momentum space ${\mathrm{Vol}}(p_y)$ in the extra dimensions is also finite
\begin{eqnarray}
{\mathrm{Vol}}(p_y) = \Omega_d \frac{L_{\mathrm f}^d}{(2 \pi)^d}  \quad, 
\end{eqnarray}
where we have assumed that in the limit of small $R$ the KK-modes have smooth spacing in
the directions of the extra dimensions.
Now consider the expansion of the wave-function $\phi$ in terms of eigenfunctions 
$\vert k \rangle = \vert p(k) \rangle$
\begin{eqnarray} 
\vert k \rangle = e^{{\mathrm i}(k_x x + k_y y) } \quad,
\end{eqnarray}
Where the wave-vector in direction of the extra dimensions $k_y$ is geometrically quantized in
steps $n/R$. The expansion then reads
\begin{eqnarray} 
\phi(x,y) = \intsum \frac{{\mathrm d}^3 k_x}{(2 \pi)^{d+3}} \frac{e^{{\mathrm i}(k_x x + k_y y) }}{N} \quad,
\end{eqnarray}
where $N$ is the normalization factor which has to be correctly set in the presence of a minimal
length. The eigenfunctions are normalized to
\begin{eqnarray} \label{norm}
\langle p'(k') \vert p(k) \rangle &=& (2 \pi )^{3+d} \delta(k'_x-k_x) \delta_{k_y'k_y} R^d \nonumber \\
&=& (2 \pi )^{3+d}  \delta(p'_x-p_x) \Bigg| \frac{\partial p_i}{\partial k_j} \Bigg| \delta_{p_y'p_y}R^d
\quad,
\end{eqnarray}
where the functional determinant of the relation is responsible for an extra factor accompanying the
$\delta$-functions. When taking the continuum limit of (\ref{norm}) we find with $\delta_{k_y'k_y} R^d \to \delta(k_y'-k_y)$ the usual normalization.

So the completeness relation of the modes takes the form
\begin{eqnarray} 
\intsum \frac{{\mathrm d}^3 k_x}{(2 \pi)^{d+3}} \frac{\langle k' \vert k \rangle}{N} = 
  R^d {\mathrm{Vol}}_d(p_y) 
\quad.
\end{eqnarray}
To avoid a new normalization of 
the eigenfunctions we take the factors into the integral by a redefinition of the measure in momentum space 
\begin{eqnarray} \label{rescalevolume3}
{{\mathrm d}^{d+3} k} \rightarrow {{\mathrm d}^{d+3} p}  \Bigg| 
\frac{\partial k_i}{\partial p_j} 
\Bigg| \frac{1}{{\mathrm{Vol}}_d(p_y) R^d} \quad.
\end{eqnarray}
This redefinition has a physical interpretation because we expect the momentum 
space to be squeezed at high momentum values and weighted less. In the standard scenario with
a non-compact momentum space we have $(2 \pi)^d {\mathrm{Vol}}_d(p_y) = {\mathrm{Vol}}_d(y)$ and thus
the factor cancels
to one.

\subsection{Minimal Length in Quantum Field Theory}

To proceed towards quantum field theory we could now take the continuum limit of (\ref{CommXP}). 
The purpose of our computations is to express all quantities in terms of the momentum $p$ as we eventually
wish to describe physical observables. Keeping the relations with the wave vector $k$ gives back the familiar
relations but does not allow us to connect to particle physics. 
However, in intermediate steps we can stick to the $k$-formalism and proceed with a minimum of modifications. 
Regarding the fact that we have to give up an easy transformation from coordinate space to momentum space we
go on with the wave-vectors and can apply Fourier transformations.  

When using the Feynman rules in $k$-space we first have to make sure that we use the right conservation law. 
As the relation between the wave-vector and the momentum is no longer linear, $k$ is not additive and it is not 
conserved in particle interactions although it is conserved for one 
propagating particle (since it is a function of a conserved quantity). So, the right conservation factor 
for the vertices with in- and outgoing momenta $p^{\rm n}$, where 'n' labels the participating particles, 
and $p^{\mathrm{tot}}_\alpha=\sum_{\rm n} p^{\rm n}_\alpha$ the total sum of the momenta is 
\begin{eqnarray} 
\delta^{4+d}\left(k(p^{\mathrm{tot}}_\alpha)\right)=\delta^{4+d}(p^{\mathrm{tot}}_\alpha) \Bigg| 
\frac{\partial p_\nu}{\partial k_\mu} 
\Bigg|. 
\end{eqnarray}

Now what about the dynamics of the particle? E.g. the Lagrangian ${\cal L}^{\phi}$ for a scalar field $\phi$
is derived by quantization of the energy momentum relation. So, we find in the continuous case
\begin{eqnarray}
{\cal L}^{\phi} = \int {\mathrm d}^{d+4}x \; \phi \left( \hat{p}(k)^2 -m^2 \right) \phi \quad.
\end{eqnarray}
As before, the modification arises solely by the fact that $\hat{p}$ is now a function of $k$.  
The propagator can then be found in $k$-space by a Fourier transformation
\begin{eqnarray}
\Delta^{\phi}(x) &=& \int d^{d+4} k \frac{e^{-ikx}}{p(k)^2-m^2} \quad,
\end{eqnarray}
and so
\begin{eqnarray}
\Delta^{\phi}(k) &=& \frac{1}{p(k)^2-m^2} \quad.
\end{eqnarray}
As is well known, the Lagrangian in the given form leads to complications in the generating functional.
Working in Minkowski-space, the path integral does not converge as the exponent, given by ${\cal L}$ is not
positive definite. We adopt the usual procedure for this problem by performing a Wick-rotation and changing
to Euclidean space. In this case, the propagator takes the form
\begin{eqnarray}
\Delta^{\phi}_{\mathrm e}(k) &=& \frac{1}{p_{\mathrm e}^2+m^2} \quad.
\end{eqnarray}
Similar derivations as for the scalar field apply for fermion fields and yield
\begin{eqnarray}
\Delta^F(k) &=& \frac{1}{p\hspace{-1.5mm}/(k)-m} \quad.
\end{eqnarray}
As expected, the propagator in $k$-space can in general be found by
the replacement $k \to p(k)$. To derive the interaction terms one has to couple gauge fields to
the free Lagrangian. It has been shown in \cite{Hossenfelder:2003jz} that in an approximation in first order (first order
as well in the couplings as in $M_{\mathrm f}$ or mixtures of both) the vertices are not modified. 

To summarize we have then the following procedure to compute diagrams:

\begin{itemize}
\item Make computations in $k$-space and apply usual Feynman rules 
\item Take the propagator as a function of $p(k)$ 
\item Use conservation of momentum on the vertices $\delta(k(\sum p))$ 
\item Finally replace the $k$-integration via 
\begin{eqnarray} \label{rescalevolume4}
\frac{{\mathrm d}^{d+4} k}{(2 \pi)^{4+d}} \rightarrow \frac{{\mathrm d}^{d+4} p}{(2 \pi)^{4+d}}  
\Bigg| \frac{\partial k_{\mu}}{\partial p_{\nu}} 
\Bigg| \frac{1}{{\mathrm{Vol}}_d(p_y) R^d} \quad.
\end{eqnarray}
\end{itemize}

\section{Minimal Length and Running Gauge Couplings}

The aim of our calculations is an investigation of the running of the gauge couplings in an energy range
$p \sim M_{\mathrm f}$. In the following, we will use the specific relation for $p(k)$ by choosing for 
the scalar function in 
 (\ref{defxi})
\begin{eqnarray}
\xi(p_{\rm e}) &=& \int_0^{p_{\rm e}} \exp\left(-\epsilon \frac{\pi}{4} p_{\rm e}^2\right) \label{model} \quad,
\end{eqnarray}
where the factor $\pi / 4$ is included to ensure, that the limiting value is $L_{\mathrm f}$. A frequently used relation
in the literature \cite{tanh} is $\xi(p)=\tanh^{1/\gamma}(p^\gamma)$, with $\gamma$ being some positive integer. 
These both choices for modeling the minimal length are compared in Fig. \ref{fig5}. As can be seen, in the considered 
energy range, the differences are negligible. The model dependence at 
smaller energies will be addressed in the
next section.

The Jacobian determinate of the function $k(p)$ is best computed by adopting spherical coordinates and can be
approximated for $p \sim M_{\mathrm f}$ with
\begin{eqnarray}
\Bigg| \frac{\partial k_{\mu}}{\partial p_{\nu}} 
\Bigg| 
&\approx& \exp\left(-\epsilon \frac{\pi}{4} p_{\rm e}^2\right) \quad.
\end{eqnarray}

Since this factor occurs as a modification to the measure in momentum space, we see clearly that
the minimal length acts essentially as a cut-off regulator. However, in difference to cut-off calculations 
in quantum field theory, here the cut-off has a physical interpretation and is cause for effects on its own. The
regulator itself is a parameter of the model. It is
the existence of a fundamental length which implies that processes involving high energies
will be suppressed and the UV-behavior of the theory will be improved. So, we are able to perform
an integration over the whole KK-tower instead of truncating the high end.

As an example we have computed the one-loop correction to the photon propagator, using the above derived
steps. This may be found in Appendix A.

The effect of the minimal length on the integration over momentum space is essentially that
the contributions at high momenta get suppressed and the loop-results with high external momenta
approach a constant value. We have two effects working against each other. On the one hand, we have
the power law arising from the extra dimensions, on the other hand we have the exponential suppression
arising from the minimal length.

The relation between the higher dimensional coupling constant $\tilde{g}_i$ 
and the four-dimensional coupling $g^2_i=4 \pi \alpha_i$ is given by the volume of the extra dimensions
\begin{eqnarray} \label{couplings4d}
g_i &=& \tilde{g}_i {\mathrm{Vol}}_d(y) \quad.
\end{eqnarray}

\begin{figure}

\includegraphics{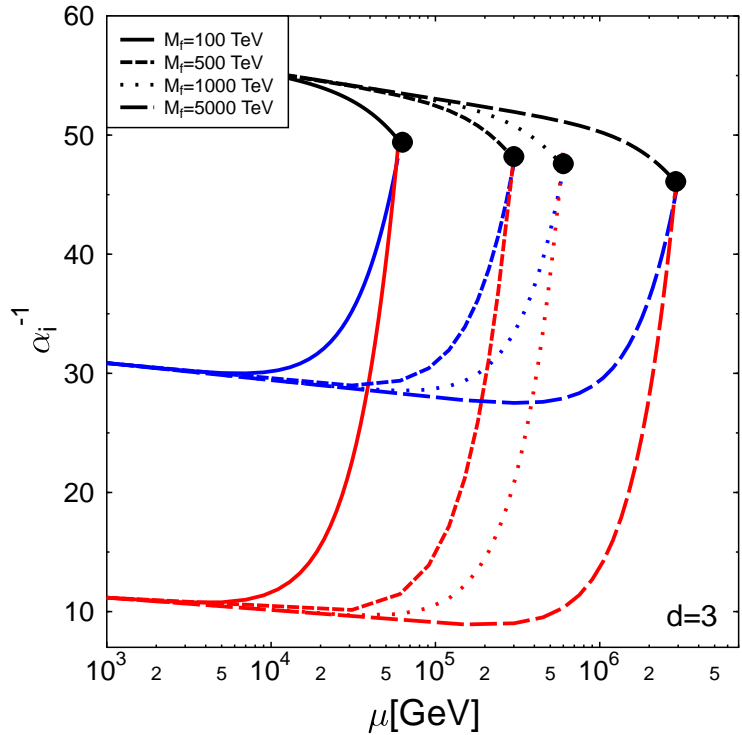}

\caption{The result for the running of the gauge couplings for $M_{\rm f}=100,500,1000,5000$~TeV and fixed 
$d=3$.
\label{fig1}}

\end{figure}

\begin{figure}

\includegraphics{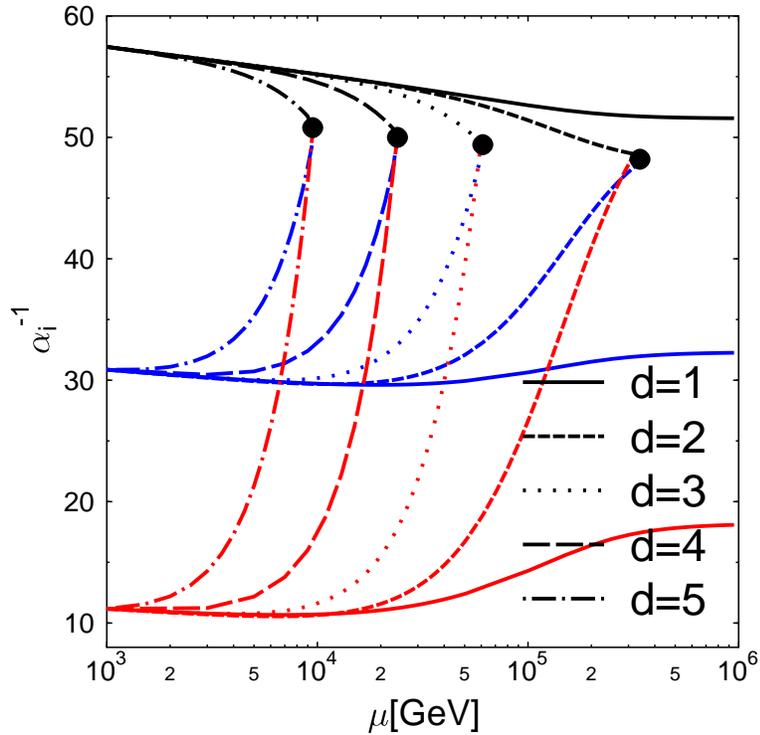}

\caption{The result for the running of the gauge couplings for $d=1,2,3,4$ and fixed 
$M_{\rm f} =100$~TeV. 
\label{fig2}}

\end{figure}
\begin{figure}

\hspace*{-0.8cm}\includegraphics{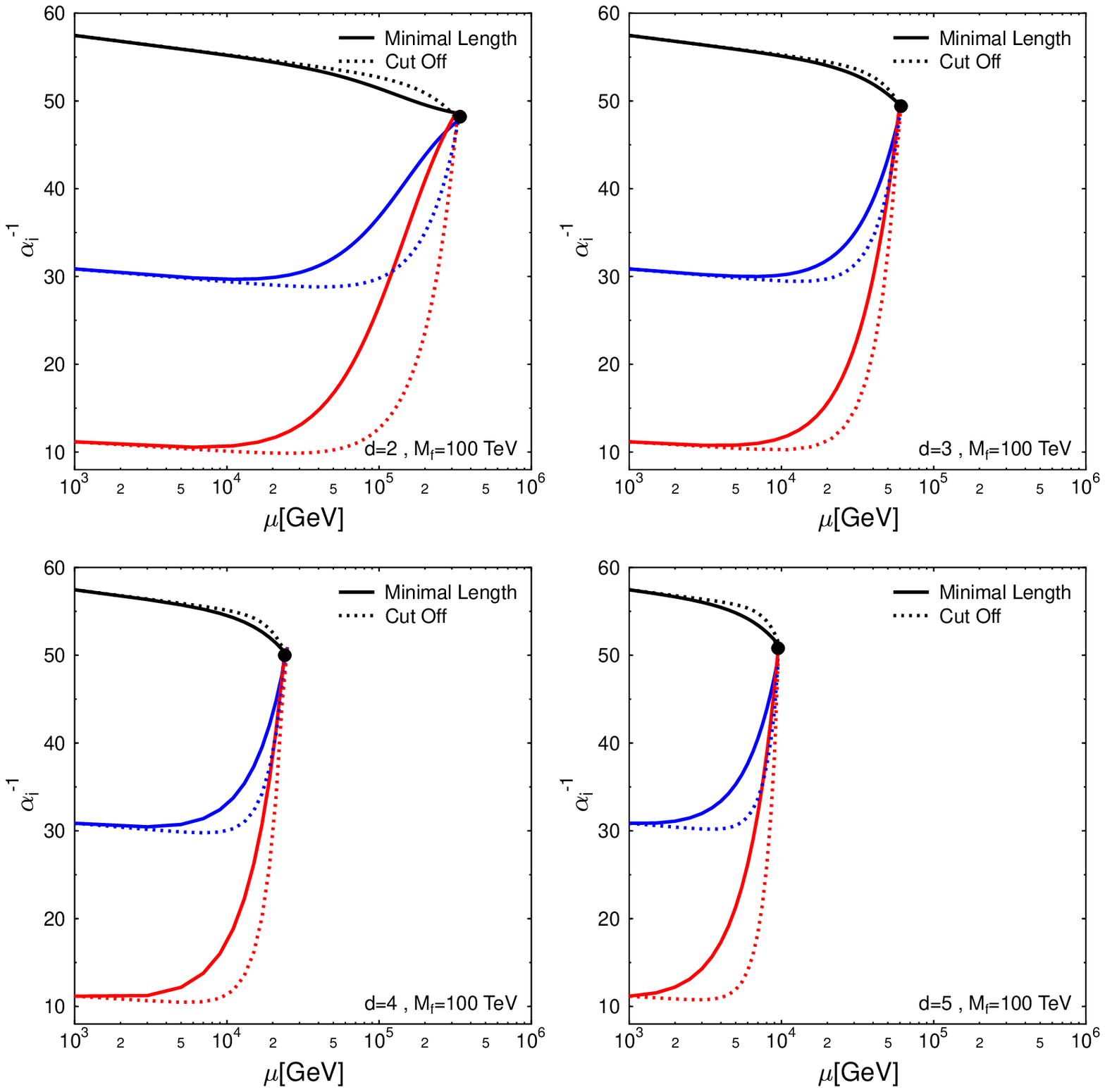}

\caption{$d=2,3,4,5$, $M_{\rm f}=100$~TeV, $M_{{\rm SUSY}}=1$~ TeV. The dotted lines show the 
 result with a hard cut-off, the solid lines the result from the minimal length. 
\label{fig3}}

\end{figure}

To examine
the running of the coupling constants $\alpha_i$, we assume that above the supersymmetry breaking scale 
$M_{\rm SUSY}$ we are dealing with the minimal supersymmetric extension of Standard Model ({\sc MSSM}),
whereas below $M_{\rm SUSY}$ we have the symmetry groups of the Standard Model.

The summarized one-loop contributions arising from the structure constants groups of the SM (after 
inclusion of the factor $3/5$ for $\alpha_1$) read
\begin{eqnarray}
{b^{\rm SM}}=[b^{\rm SM}_1, b^{\rm SM}_2, b^{\rm SM}_3]&=&[4,10/3,-7] \quad.
\end{eqnarray}  

Within the {\sc MSSM}, the number of fermion generations $n_{\rm g}=3$ and
the number of Higgs-fields $n_{\rm h}=2$ we have then above $M_{\rm SUSY}$ the coefficients
\begin{eqnarray}
b=[b_1,b_2,b_3]&=&[0,-6,-9]+n_{\rm g}[2,2,2]+n_{\rm h}[3/10,1/2,0]\nonumber\\&=&[33/5,1,-3]\quad.
\end{eqnarray} 
As pointed out in\cite{Dienes:1998vh} these supersymmetric $b_i$ coefficients will change in a higher
dimensional spacetime due to the different content of the superfields. This content of the KK-excitations of the fields 
can be accommodated in hyper multiplets of $N=2$ super symmetry instead of the $N=1$ super symmetry 
in the four dimensional spacetime. Therefore, the modified one-loop contributions have factors 
different from the {\sc MSSM} ones. In this paper, we will consider only the case in which all fermions 
are confined to the brane ($n_{\rm g}=0$). Then the factors for the excitation modes are given by
\begin{eqnarray}
\tilde{b}=[\tilde{b}_1,¸\tilde{b}_2,\tilde{b}_3]&=&[0,-4,-6]+n_{\rm h}[3/10,1/2,0] \quad.
\end{eqnarray}  
The running of the couplings  above the scale of {\sc SUSY}-breaking $M_{\rm SUSY}$ is given by the 
familiar expression
\begin{eqnarray} \label{alphadurchalpha}
\frac{\tilde{\alpha}_i(q')}{\tilde{\alpha}_i(q)} &=& 1-(b_i-\tilde{b}_i)
\left[\pi(q,0)-\pi(q',0)\right]-\tilde{b}_i \left[ \pi(q,d)-\pi(q',d)\right] \quad,
\end{eqnarray} 
where $\pi(q,d)$ denotes the finite part of the scalar factor in the one-loop contribution, which leads to
a renormalization of the gauge-field propagator. It should be noted, that the inclusion of the minimal
length does not remove infrared divergences. Thus, a proper regularization is still necessary, resulting in
a difference between 'bare' and 'physical' couplings.

The higher dimensional one-loop contributions to the propagator can now be calculated
by using the formalism developed 
in section 4. We find that the  infrared regularized result can be given in the integral form (see Appendix A)
\begin{eqnarray} \label{piqd}
\pi(q,d) &=&  3 b_i  \frac{\alpha_i}{2 \pi}  \frac{(2 \pi)^{d}}{\Omega_d} (\pi \tilde{\epsilon})^{d/2} \left[
  \int_0^{1} {\rm d}x  x (1-x)^{1+d/2} \int_{\tilde{\epsilon}}^{\infty} \hspace{-3mm}{\rm d}z \;
 e^{\displaystyle - z x q^2}  {z} ^{-1-d/2} \right. \nonumber \\
 &+& \left. \frac{1}{q^2} \frac{(d+4)}{2(d+3)} {\tilde{\epsilon}}^{-1-d/2} \int_0^{1} {\rm d}x  (1-x)^{1+d/2}  \;
 \left( e^{\displaystyle - \tilde{\epsilon} x q^2} -1 \right) \right] \quad,
\end{eqnarray}
with the abbreviation $\tilde{\epsilon}=\epsilon\pi/4$. The result does depend explicitly on the 
parameter $\epsilon$ since this is a physical quantity in our description. As expected, we find two
effects: the first giving a power law behavior (the power depending on d) which can be located in
the power of $z$, the second an exponential drop
due to the minimal length, which can be located in the non-zero lower bound of $z$-integration.

Let us briefly compare this with the result using the hard cut-off computation where the 
sliding scale $q$ is identified with
the cut-off $\Lambda$ (see e.g. \cite{Dienes:1998vh}). It is obvious, that in our scenario the role
of the UV cut-off is given to $L_{\mathrm f}$. We thus interpret the only free parameter as energy scale:
\begin{eqnarray}  \label{hardcutoff}
\frac{\tilde{\alpha}_i(q')}{\tilde{\alpha}_i(q)} = 
1 &-& \alpha_i(q') \frac{b_i}{2\pi} \ln \frac{q}{q'} \nonumber \\ &+&  
\Theta(q-\mu_0) \alpha_i(q') \frac{\tilde{b}_i}{2\pi}  \left( \ln \frac{q}{\mu_0} 
-   \frac{\Omega_d}{d} \chi_d^d \left[ \left( q L_{\mathrm f} \right)^d - \left( \mu_0 L_{\mathrm f}) \right)^d\right] \right) \quad.
\end{eqnarray}
Here $\mu_0$ is the matching point below which the four-dimensional logarithmic running is unmodified and $\chi_d$ is an
unknown factor usually set to be equal one. In the above expression, $\Theta$ denotes the Heaviside-function. 

The comparison to our result is best done when making a power series expansion of the integral form  (\ref{piqd}) for small $\epsilon$. 
For $ \Delta \pi(q,q',d) = \pi(q,d)-\pi(q',d)$
we find
\begin{eqnarray} \label{series0}
 \Delta \pi(q,q',0) &=& b_i   \frac{\alpha_i}{2 \pi} \left[ \ln \left(\frac{q}{q'}\right) - 
\frac{1}{3} \tilde{\epsilon}\left( q^2-q'^2 \right) + {\cal{O}}(\tilde{\epsilon}^2) \right] \\
 \Delta \pi(q,q',1) &=& b_i   \frac{\alpha_i}{2 \pi} \left[\frac{9 }{32} \pi^3 \tilde{\epsilon}^{1/2} 
\left( q-q' \right) - \frac{74  }{105} \pi^{3/2} \tilde{\epsilon}\left( q^2-q'^2 \right)  
+ {\cal{O}}(\tilde{\epsilon}^2) \right]\\
 \Delta \pi(q,q',2) &=& b_i   \frac{\alpha_i}{2 \pi} \left[-\frac{2 }{5}\pi^{3}  \tilde{\epsilon}\left( 
 q^2 \ln \tilde{\epsilon}q^2 - q'^2 \ln \tilde{\epsilon}q'^2 \right) - \right.\nonumber\\
&&\left. \;\; \quad   \frac{ \pi^{3}}{150} \tilde{\epsilon}
(60 \gamma -89)\left( q^2-q'^2 \right)  + {\cal{O}}(\tilde{\epsilon}^2)  \right]\\
 \Delta \pi(q,q',3) &=& b_i   \frac{\alpha_i}{2 \pi} \left[ \frac{656}{393} \pi^{9/2} \tilde{\epsilon}
\left( q^2-q'^2 \right)  - \frac{5  }{32} \pi^{6}\tilde{\epsilon}^{3/2}\left( q^3-q'^3 \right) +\right.\nonumber\\
&&\left. \;\; \quad   \frac{2528 }{9009} \pi^{9/2} \tilde{\epsilon}^2\left( q^4-q'^4 \right) +   
{\cal{O}}(\tilde{\epsilon}^3)  \right]\\
 \Delta \pi(q,q',4) &=& b_i   \frac{\alpha_i}{2 \pi}\left[ - \frac{4}{7} \pi^6 \tilde{\epsilon}(q^2-q'^2) - \frac{869-420 \gamma}{2450}\pi^6 \tilde{\epsilon}^2 (q^4-q'^4) + \right. \nonumber\\
&&\left. \;\; \quad  \frac{42}{245} \tilde{\epsilon}^2 \pi^6 \left(q^4 \ln \tilde{\epsilon}q^2 - q'^4
\ln\tilde{\epsilon}q'^2\right) + {\cal{O}}(\tilde{\epsilon}^3)  \right]\\
 \Delta \pi(q,q',5) &=& b_i   \frac{\alpha_i}{2 \pi}  \left[ \frac{592}{1287} \pi^{15/2} 
\tilde{\epsilon} \left( q^2-q'^2 \right) - \right.\nonumber \\
&& \quad \quad \frac{928}{2145}\pi^{15/2} \tilde{\epsilon}^2\left( q^4-q'^4 \right)+  
\frac{7}{128}\pi^9 \tilde{\epsilon}^{5/2}\left( q^5-q'^5 \right) - \nonumber\\
&&\left. \;\; \quad  \frac{8768}{109395} 
  \pi^{15/2} \tilde{\epsilon}^3\left( q^6-q'^6 \right)+ 
 {\cal{O}}(\tilde{\epsilon}^4)   \right] \label{series5}\quad,
\end{eqnarray}
For $d=0$ we find the familiar 
logarithmic divergence. For higher $d$ we find that an odd number of
extra dimensions leads to one-loop corrections with a power law, whereas for an
even number of extra dimensions there is a mixture of the $d$-power term with a 
logarithmic contribution. It can be seen, that in contradiction to the results from introduction of a cut-off in
momentum space, the leading power is not $d$. This conclusion agrees with analyzes from \cite{oliver} using
dimensional regularization. It is interesting to note, that in the limit $R\gg L_{\mathrm f}$ the result does 
no longer depend on the value of the radius of the extra dimensions.

The scale $\mu_0$ in  (\ref{hardcutoff}) usually is chosen to be $1/R$. This yields a good agreement with our 
minimal length scenario  for $1/R$ close to $M_{\rm SUSY}$ and particular values of $\chi_d$. However, 
for even values of $d$ the power law in (\ref{hardcutoff}) is not a good fit. 

There are three main points which are new to our results: 
\begin{itemize}
\item Using the minimal length we do not need to introduce an initial threshold (in addition to the symmetry breaking
scale) as we can include {\sl all} virtual KK-excitations.
\item There is no arbitrariness for the parameter $\chi_d$ and the identification of the energy scale.
\item The couplings do no longer run with a pure power law.
\end{itemize}

\section{Numerical Results}

In the following we will compare the full result  (\ref{piqd}) to the cut-off result and give
numerical values for $\chi_d$ in the parametrization  (\ref{hardcutoff}). This numerical fit is optimized
to best reproduce the unification point of the full result. We will set $\mu_0=M_{\rm SUSY}=1/R$ and match the curve with
the Standard Model result at this energy.

For the initial values we use the data set\cite{particledata}
\begin{eqnarray*}
M_Z &=& 91.197 \pm 0.007 {\rm GeV}\\
\alpha_1\left(M_Z\right)^{-1} &=& 58.98 \pm 0.04 \\
\alpha_2\left(M_Z\right)^{-1} &=& 29.57 \pm 0.03\\
\alpha_3\left(M_Z\right)^{-1} &=& 8.5 \pm 0.5 \quad.
\end{eqnarray*} 

In Figure \ref{fig1} the result of our computation for fixed $d=3$ and different values of $L_{\mathrm f}$
is shown. 
We see that the onset of the
deviations from the 4-dimensional result is roughly given by the inverse minimal length and the unification
point lies at an energy scale of the same order of magnitude. The 
value of the coupling at the unification point does not vary much and lies at $1/\alpha_i \approx 50$.

Fig. \ref{fig2} shows the results of our computation for fixed $M_{\mathrm f}=$~100 TeV and different values of $d$. 
Here it shows clearly how the two factors -- the power law running and the dumping from the minimal length -- 
act 
against each other. For $d=1$ the minimal length avoids unification. For $d>1$ it can be seen, that a higher 
$d$ leads to a faster running and the unification point is reached before the exponential suppression becomes
important.

\begin{figure}

\includegraphics{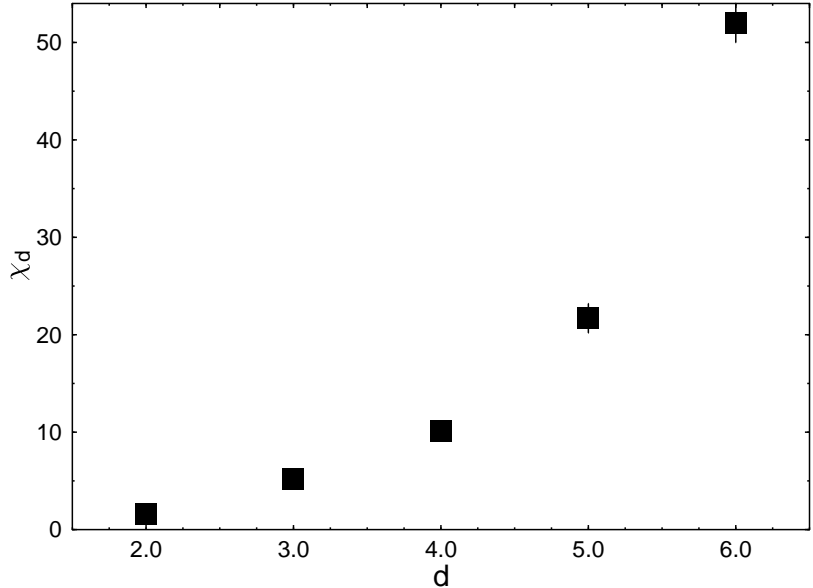}

\caption{The values of the fitting parameter $\chi_d$. $\chi_2 = 2,1 \pm 0.1$, 
$\chi_3 = 5.2 \pm 0.2$,
$\chi_4 = 10.1 \pm 0.4$
$\chi_5 = 21.7 \pm 0.8$
$\chi_6 = 52 \pm 1.9$ 
\label{fig4}}

\end{figure}

Fig. \ref{fig3} shows a comparison of our result with the cut-off result, using the fitting parameter $\chi_d$,
whose values are depicted in Fig. \ref{fig4}. The errors are mainly due to the fact that in all cases the
unification does not occur at one exact point. 

\begin{figure}

\includegraphics{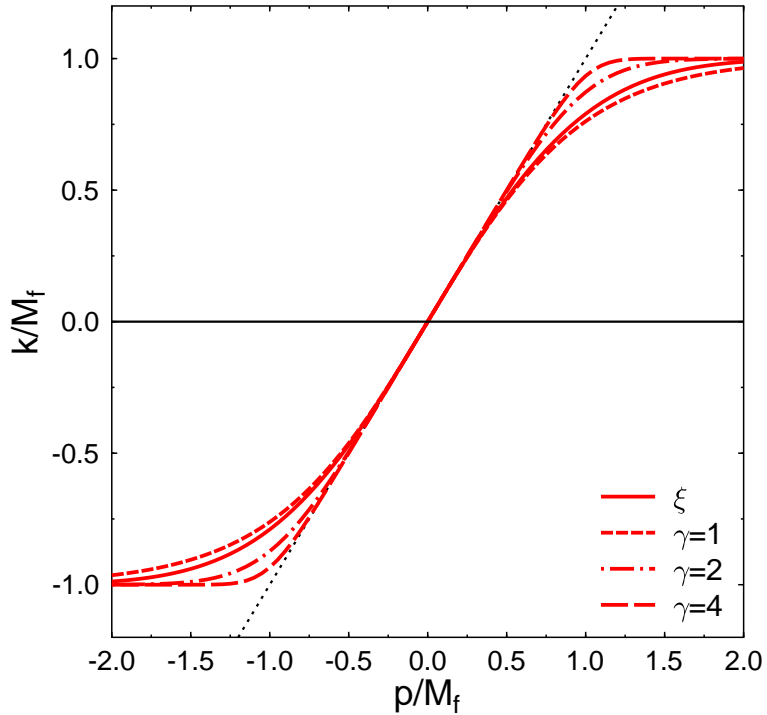}

\caption{The linear dotted line shows the case of no modification $k=p$. 
\label{fig5}}

\end{figure}

Note, that 
our specific choice of the functional relation, although not relevant for qualitative statements, 
introduces an additional model dependence at $p < M_{\mathrm f}$. To parametrize the lack of knowledge
about the exact relation $k(p)$, consider the expansion
\begin{eqnarray}
\xi (p_{\mathrm e}) &=& \sum_{i=0}^{n} c_i \left( \frac{p_{\mathrm e}}{M_{\mathrm f}} \right)^n 
e^{\displaystyle -\epsilon p_{\mathrm e}^2} \quad,
\end{eqnarray}
with $c_0=1$. 
The parameters in this series can be transformed into parameters in the functional determinant and
further into parameters in the final expansion (\ref{series0}) - (\ref{series5}). The running of the
coupling in this energy range therefore leads a direct connection to the behavior of the minimal
length.
The plot in Fig. \ref{fig5} shows a comparison of different relations for $k(p)$. The dashed lines depict the
function $\tanh^{1/\gamma}(p^\gamma)$ for different values of $\gamma$. The solid line beetween them shows
our relation $\xi(p)$. 

Further we want to note, that the above used assumption $R \gg L_{\mathrm f}$ which justifies the replacement
of the KK-sum with an integral, leads numerically quiet good results even in the region, where $R$ and
$L_{\mathrm f}$ differ only by one order of magnitude. The approximation however, breaks down for 
$R\to L_{\mathrm f}$ as in
this case the minimal length would avoid the existence of excitations at all.

\section{Conclusion}

In this paper we computed the running of the gauge couplings in a higher dimensional space 
time at one loop order. We proposed to remove the UV-divergences with the introduction 
of a minimal length scale and examined the results on their dependence of the parameters.
We found that the minimal length acts as a natural regulator. The scale dependence of the gauge 
couplings revealed a powerlaw at energies below the inverse minimal length and stagnated at energies
much higher than the inverse minimal length. In this high energy region, the generalized 
uncertainty principle
does not allow a further resolution of structures. The derived result for $d>1$ confirmes the cut-off 
regularized result and enriches the regularization scheme with a physical interpretation.

\section*{Acknowledgments}
I would like to thank Keith Dienes for valuable discussions and his contribution to this work.
Further, I want to thank Stefan Hofmann and J\"org Ruppert for their answers as well as 
for their questions. This work
was supported by a fellowship within the Postdoc-Programme of the German Academic 
Exchange Service 
({\sc DAAD}) and NSF PHY/0301998.

\section*{Appendix A}

As an example we compute the {\sc QED} one-loop contribution to the photon propagator under inclusion
of the modifications arising from the generalized uncertainty principle. The photon carries the
external momentum $q$ and therefore propagates on the brane. Here, we
will treat the fermions circling in the loop as a higher dimensional particle, even 
if we do not consider this case in the context of this paper.
The result for loops of gauge-bosons, which are allowed to leave the brane, is similar except for 
a constant factor arising from the structure constants of the gauge group. In the familiar way, 
all contributions can finally be summarized in the $b_i$-coefficients. 

Since the mass of the fermions is negligible at the energy scales $\approx M_{\mathrm f}$ that we are interested in,
we treat the particle as massless. Throughout this Appendix we perform the calculation in Euclidean space
and suppress the index 'e'.

With the abbreviation $\tilde{\epsilon}= \epsilon \pi/4  $ the Feynman rules give as explained in the text 
\begin{eqnarray} \label{selfenergy1}
\Pi_{\mu\nu}(q,d)= {e}^2 
  \frac{(2 \pi)^{d}}{\Omega_d} \tilde{\epsilon}^{d/2}  \int \frac{d^{4+d} p}{(2 \pi)^4} 
{\mathrm{\large Tr}}
\left[\frac{\gamma_{\mu}}{p\hspace{-1.5mm}/} \right] \left[
\frac{\gamma_{\nu}}{p\hspace{-1.5mm}/-q\hspace{-1.5mm}/} \right] e^{\displaystyle -\tilde{\epsilon}  p^2} \quad,
\end{eqnarray}
where the above expression is understood to result after the Wick-rotation, and where we have replaced the sum over 
KK-modes by an approximate integral. 
We thus perform a higher 
dimensional computation instead of using the effective theory on our brane. Since the external momentum $q$
lies on our brane, it does not mix with the internal momenta $p$ and in an effective description the excitations
therefore appear as a tower of massive particles. This effective theory on the brane is completely equivalent 
to the above one in the whole bulk.

As explained in
the text, the zero mode needs further treatment because the $b_i$ factors are different when lying 
on the brane only. This is taken into account with the 2nd factor in  (\ref{alphadurchalpha}) 
using the coefficients
$\tilde{b}_i$. The zero mode is included in the above integral but with the wrong factor from 
the bulk modes. It
therefore has to be subtracted and replaced with the brane-only term as in\cite{Dienes:1998vh}.

It should be noted, that the above expression is gauge invariant as the formalism developed respects
all symmetries in Euclidean space. To see this contract the above expression with $q$. Gauge invariance
then demands $q^\mu \Pi_{\mu \nu}=0$. This can be written as
\begin{eqnarray} \label{gauge1}
q^\mu\Pi_{\mu\nu}(q,d) &\propto&  \int \frac{d^{4+d} p}{(2 \pi)^{4+d}} \;
{\mathrm{\large Tr}}
\left[\frac{q\hspace{-1.5mm}/}{p\hspace{-1.5mm}/} \right] \left[
\frac{\gamma_{\nu}}{p\hspace{-1.5mm}/-q\hspace{-1.5mm}/} \right] e^{\displaystyle -\tilde{\epsilon}  p^2} \quad.
\end{eqnarray}
Now we rewrite expression and return back to $k$-space to find
\begin{eqnarray} \label{gauge2}
q^\mu\Pi_{\mu\nu}(q,d) &\propto&  \int \frac{d^{4+d} k}{(2 \pi)^{4+d}} \; 
{\mathrm{\large Tr}}  
\left[
\frac{1}{p\hspace{-1.5mm}/(k)-q\hspace{-1.5mm}/} - \frac{1}{p\hspace{-1.5mm}/(k)} \right] \gamma_{\nu} \quad.
\end{eqnarray}
Now we note that substituting $p' \to p-q$ in the first term does not modify the contours of integration as the
asymptotic value of $k(p')$ is still $M_f$. So the two terms are identical and cancel, keeping gauge invariance.

We then can assume  
\begin{eqnarray} \label{defpiPi}
\Pi_{\mu\nu}(q,d)=\pi(q,d)(q_{\mu}q_{\nu}-g_{\mu\nu}q^2) \quad.
\end{eqnarray}
By taking the trace of  (\ref{selfenergy1}) and using  (\ref{defpiPi}) we find\footnote{As mentioned in 
\cite{oliver} the trace over
the higher dimensional $\gamma$-matrices yields an unwanted factor $2^d$. This is due to the 
compactification scheme which
is unsuitable for fermions as it does not properly reproduce the degrees of freedom on the brane. 
We too therefore drop 
this factor by hand.}
\begin{eqnarray} \label{selfenergy2}
\pi(q,d)=  \frac{ {e}^2}{q^2} \frac{4(2+d)}{(3+d)} 
\frac{(2 \pi)^{d}}{\Omega_d} \tilde{\epsilon}^{d/2}  \int \frac{d^{4+d} p}{(2 \pi)^4} 
\frac{p^2-pq}{p^2(p-q)^2}  e^{\displaystyle -\tilde{\epsilon}  p^2} \quad,
\end{eqnarray}

Using a modified version of the Schwinger Proper time formalism 
\begin{eqnarray} \label{eps}
\frac{e^{- \displaystyle \tilde{\epsilon} p^2} }{p^2} = -  
\int_{\tilde{\epsilon}}^{\infty  } \hspace{-3mm}{\rm d}z \;
e^{- \displaystyle z p^2}  \quad,
\end{eqnarray}
as well as the usual one with $\tilde{\epsilon}=0$
we can further simplify the integral. At this stage it is apparent why the Euclidean norm is essential since the
expression on the rightside in (\ref{eps}) otherwise would not converge. 

We then arrive at
\begin{eqnarray} \label{selfenergy3}
\pi(q,d) &=&  \frac{ {e}^2}{q^2} \frac{4(2+d)}{(3+d)} \frac{(2 \pi)^{d}}{\Omega_d} \tilde{\epsilon}^{d/2} 
\times \nonumber\\
&&    \int \frac{d^{4+d} p_e}{(2 \pi)^4} 
  \int_0^{\infty} \hspace{-3mm}{\rm d}z_1 \int_{\tilde{\epsilon}}^{\infty} \hspace{-3mm}{\rm d}z_2 \;
(p^2-pq)\;  e^{\displaystyle - z_1 (p-q)^2 - z_2 p^2} \quad.
\end{eqnarray}
After substituting $l:=p-q z_1/(z_1+z_2)$ and interchange of the $z_i$ with the momentum integral, we can
perform the momentum integration using the identities
\begin{eqnarray} \label{gauss}
\int  {\rm d}^n x \; e^{\displaystyle - a x^2} &=& \left( \frac{\pi}{a}\right)^{n/2} \\
\int  {\rm d}^n x \; x^2 e^{\displaystyle - a x^2} &=& \frac{n}{2a} \left( \frac{\pi}{a}\right)^{n/2}  \quad.
\end{eqnarray}
We use the further substitution $z_1 \to x:= z_1/(z_1+z_2)$ and relabel $z_2$ to $z$ in order to
allow an easy comparison to the standard result. Our expression for the one-loop correction then reads
\begin{eqnarray} \label{selfenergy4}
\pi(q,d) &=&   \frac{ \alpha }{\pi q^2} \frac{(2+d)}{(3+d)} \frac{(2 \pi)^{d}}{\Omega_d} (\pi \tilde{\epsilon})^{d/2} 
\times \nonumber\\
&&   
  \int_0^{1} {\rm d}x  \int_{\tilde{\epsilon}}^{\infty} \hspace{-3mm}{\rm d}z \;
 e^{\displaystyle - z x q^2} \left( \frac{1-x}{z}\right)^{1+d/2} \left( \frac{d+4}{2z} -x q^2\right)\quad.
\end{eqnarray}
Integrating the first term by parts yields
\begin{eqnarray} \label{selfenergy5}
\pi(q,d) &=&   3 b   \frac{\alpha}{2 \pi}  \frac{(2 \pi)^{d}}{\Omega_d} (\pi \tilde{\epsilon})^{d/2} \left[
  \int_0^{1} {\rm d}x  x (1-x)^{1+d/2} \int_{\tilde{\epsilon}}^{\infty} \hspace{-3mm}{\rm d}z \;
 e^{\displaystyle - z x q^2}  {z} ^{-1-d/2} \right. \nonumber \\
 &+& \left. \frac{1}{q^2} \frac{(d+4)}{2(d+3)} {\tilde{\epsilon}}^{-1-d/2} \int_0^{1} {\rm d}x  (1-x)^{1+d/2}  \;
 e^{\displaystyle - \tilde{\epsilon} x q^2} \right] \quad,
\end{eqnarray}
where we have identified $b=4/3$ as the beta-function coefficient of our single Dirac fermion. 
The second term in  (\ref{selfenergy5}) contains the infrared divergence. As we assume that the finite part of $\pi(q,d)$ which is
of interest for our running coupling fulfills the requirement $\pi(0,d)=0$, which is necessary to preserve
the pole-structure of the propagator, we subtract the divergent term and arrive at
\begin{eqnarray} \label{selfenergy6}
\pi(q,d) &=&  3 b   \frac{\alpha}{2 \pi} \frac{(2 \pi)^{d}}{\Omega_d} (\pi \tilde{\epsilon})^{d/2} \left[
  \int_0^{1} {\rm d}x  x (1-x)^{1+d/2} \int_{\tilde{\epsilon}}^{\infty} \hspace{-3mm}{\rm d}z \;
 e^{\displaystyle - z x q^2}  {z} ^{-1-d/2} \right. \nonumber \\
 &+& \left. \frac{1}{q^2} \frac{(d+4)}{2(d+3)} {\tilde{\epsilon}}^{-1-d/2} \int_0^{1} {\rm d}x  (1-x)^{1+d/2}  \;
 \left( e^{\displaystyle - \tilde{\epsilon} x q^2} -1 \right) \right] \quad.
\end{eqnarray}
An analytic expansion in a power series in $\tilde{\epsilon}$ reveals the differences relative to the
pure power law-running and is given in (\ref{series0})-(\ref{series5}).

\end{document}